# Bayesian Re-Analysis of the Phylogenetic Topology of Early SARS-CoV-2 Case Sequences

**Michael B. Weissman**

**Department of Physics, University of Illinois at Urbana-Champaign,**
**1110 W. Green St. 61801, USA**

**mbw@illinois.edu**

Orcid 0000-0003-0081-4429

## Abstract

A much-cited 2022 paper by Pekar et al. claimed that Bayesian analysis of the molecular phylogeny of early SARS-CoV-2 cases indicated that it was more likely that two successful introductions to humans had occurred than that just one had. Here I show that after correcting a fundamental error in Bayesian reasoning the results in that paper give larger likelihood for a single introduction than for two.

## Introduction

Pekar et al. [1] (denoted P2022 henceforth) is one of the most influential papers addressing the origins of the SARS-CoV-2 (SC2) virus. P2022 makes the qualitative claim that Bayesian analysis of the RNA sequences found in early cases indicates that they came from two successful introductions from a non-human host. Although several major coding implementation errors found by McCowan[2][3] have been corrected in a revised version, here I point out that the major remaining errors[2] in statistical analysis can be approximately corrected just using the published P2022 results.

While there are interesting questions concerning the relevance of the one-spillover vs. two-spillover issue to the origins of SC2 and concerning the appropriateness of the evolutionary model chosen by P2022, I will not address those here. I will focus on showing how fundamental errors in basic statistical methods in P2022 can be repaired with minimal auxiliary assumptions using the data in the paper.

## The Central P2022 Argument

Fig. 1, taken directly from P2022, illustrates an example of its central hypothesis. It is that SC2 had two successful spillovers to humans after circulating in some intermediate host. One of those spillovers gave rise to the successfully propagating lineage A and the other to the successfully propagating lineage B. These two lineages differ by only two nucleotides out of ~30,000 in the entire sequence. I follow the P2022 notation to call this two-introduction hypothesis "$I_2$". The competing hypothesis (denoted $I_1$) is that there was only a single successful spillover giving rise to both lineages. The "Observed Phylogeny" box on the right of Fig. 1 illustrates the $I_1$ hypothesis if its left-most sequence is taken to be the spillover sequence. In this paper I limit analysis to P2022's qualitative claim that its modeling supports "the hypothesis that lineages A and B represent separate introductions."

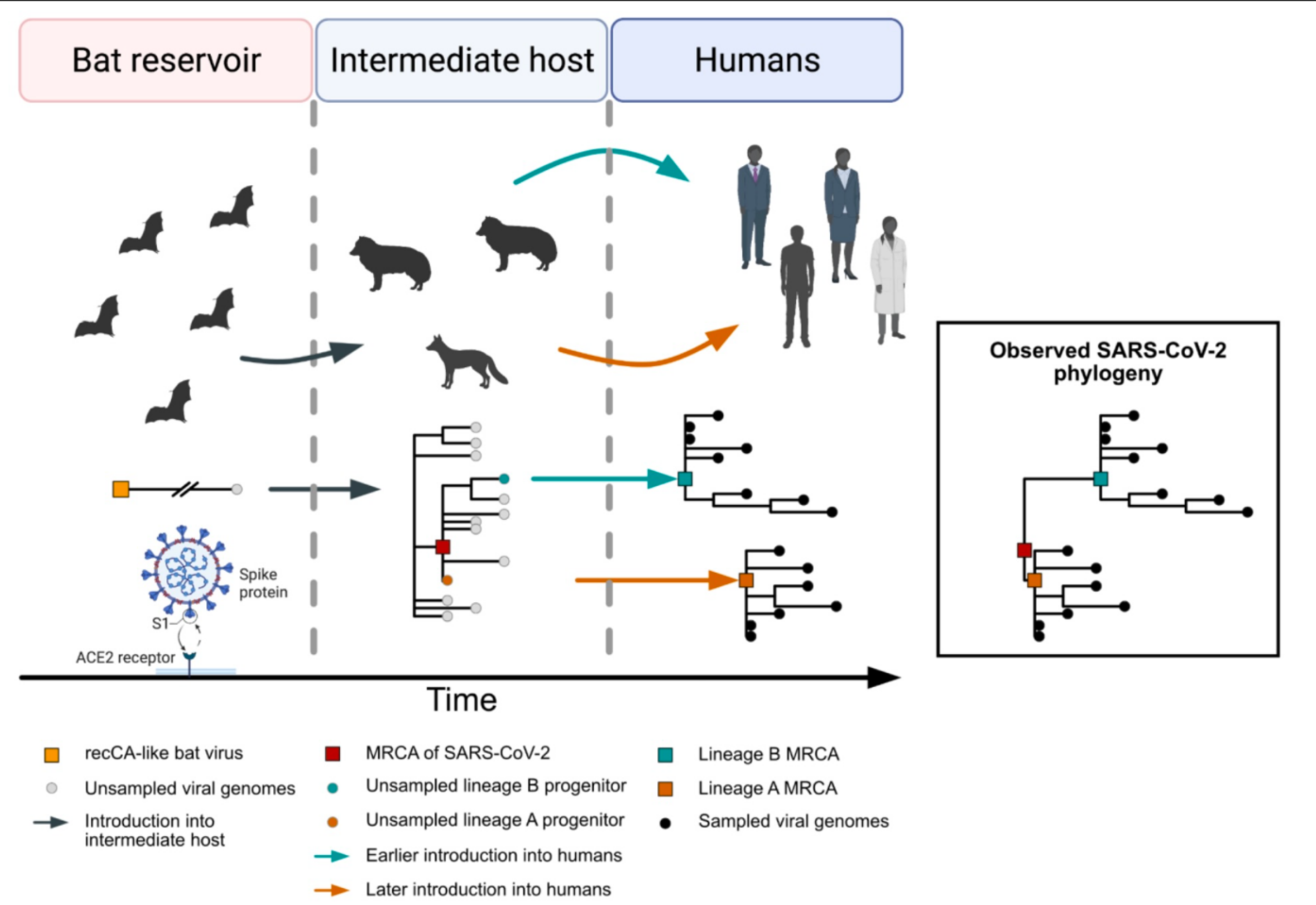

**Figure S30. Schematic depicting the multiple zoonotic origin of SARS-CoV-2.** A recCA-like virus was circulating in bats, and likely after gaining the ability to bind ACE2, jumped into an intermediate host. Therein, lineages A and B appeared and were separately introduced into humans shortly thereafter. An example phylogeny of viruses in the intermediate host is depicted, leading to separate phylogenies for lineages A and B. The resulting SARS-CoV-2 phylogeny from the combined lineage A and B viruses is presented in the black box. This scenario depicts a lineage A ancestral haplotype. See Figure S31 for intermediate and lineage B ancestral haplotypes.

Fig. 1. A pictorial guide to the type of scenarios used for the two-introduction hypothesis, reproduced from Fig. S30 of P2022, used here under the Creative Commons license https://creativecommons.org/licenses/by/4.0/.

P2022 presents a Bayesian analysis to compare the likelihoods of $I_1$ and $I_2$ for some chosen features. The standard Bayesian method to use such evidence is to compare the conditional probabilities of obtaining the observed results under the two hypotheses, i.e. compare the likelihoods of the hypotheses: P(data|$I_2$) vs. P(data|$I_1$). The odds P($I_2$)/P($I_1$) are updated by the ratio P(data|$I_2$)/P(data|$I_1$) for each new independent piece of data using Bayes theorem.

$$\frac{P_{new}(I_2)}{P_{new}(I_1)} = \frac{P_{prior}(I_2)}{P_{prior}(I_1)} \frac{P(data|I_2)}{P(data|I_1)}$$

In order to make the likelihood comparisons P2022 uses a stochastic model of how after a single introduction the virus transmits between nodes that have a range of connectivity while randomly mutating. It does not include any model of how the virus transmits and mutates before introduction into humans. That omission that leads to problems in analyzing $I_2$ since $I_2$ inherently includes a pre-introduction phase. I shall show, however, that P2022 presents information sufficient to make conservative estimates of the corrections needed.

The P2022 likelihood estimations require running many simulations of possible outputs of the model. It is impractical to model the conditional probabilities of obtaining any specific large data set of sequences because those probabilities are far too low to be picked up in a reasonable number of simulations. Instead, P2022 uses the simulations to calculate the likelihoods of the two hypotheses using some selected properties of the observed sequences.

P2022 bases its analysis on 787 sequences, obtained after some selection, from the tens of thousands of cases that occurred before Feb. 14, 2020. P2022 picks out two features of the observed phylogeny that do not seem to sit especially well with $I_1$, the single-spill hypothesis. One feature is that the sequence set includes neither of the two possible intermediate sequences on the path between A and B, i.e. sequences differing from A and from B by one nucleotide each. That would require either that both single-nucleotide mutations occurred in a single transmission step, as sometimes happens, or that the sparse sampling failed to pick up either the intermediates on the evolutionary path between A and B or their intermediate descendants. The other feature is that the two lineages have about the same size (i.e. number of cases), which might seem surprising if one lineage branched off from the other.

P2022 defines a "topology" of the sequences, which it denotes "τ", to capture the size and sequence difference features and other properties describing the different descendants of the root sequence.

> "a topology corresponding to a single introduction of an ancestral C/C haplotype—characterized by two clades, each comprising ⩾30% of the taxa, possessing a large polytomy at the base, and separated from the MRCA by one mutation was only observed in 0.0% of our simulations. Further, a topology corresponding to a single introduction of an ancestral lineage A or lineage B haplotype—characterized by a large basal polytomy and a large clade, comprising between 30 and 70% of taxa, two mutations from the root with no intermediate genomes—was observed in only 3.1% of our simulations"

To clarify, the features chosen were that:

1. Each lineage derives from a root sequence that constitutes "a large basal polytomy", i.e. at least 100 directly descendant taxa from a single root. (Other values for the minimum number were used in some auxiliary comparisons.) To illustrate, in Fig. 1 (P2022's S30) the sketch in the "Observed Phylogeny" box shows two small polytomies, one each at the roots of lineages A and B.
2. The lineage roots are compatible either with one being the probabilistically inferred most recent common ancestor (MRCA) of all the viral sequences in humans or both being one mutation downstream from that MRCA.
3. The numbers of cases in the two lineages are comparable. P2022 requires the size ratio SR between the lineages be in the interval (3/7, 7/3), i.e. $|\ln(SR)| \leq 0.85$.
(The observed SR was 0.352/0.648, i.e. $|\ln(SR)|=0.61$, so a proper Bayesian calculation would use a probability density function $PDF(|\ln(SR)|=0.61)$ but in this case the distributions of $\ln(SR)$ for both $I_1$ and $I_2$ are broad enough for the ratio $P(|\ln(SR)| \leq 0.85|I_2)/ P(|\ln(SR)| \leq 0.85|I_1)$
to be essentially the same as the ratio $PDF(|\ln(SR)|=0.61|I_2)/ PDF(|\ln(SR)|=0.61|I_1)$.)
4. The two lineages differ by 2 nucleotides, which I denote D=2.

Whenever a few properties of high-dimensional data are chosen post-hoc for statistical analysis, multiple-comparison issues arise that are not present for pre-specified analyses. Although that issue, shared by frequentist and Bayesian analyses, is relevant here, I shall dwell on a more unusual problem.

*Regardless of whether the P2022 model of post-introduction descent and ascertainment is realistic it is a specific well-defined mathematical model applied to a specific data set. Therefore it has well-defined implications. Like any mathematical implications, those can be calculated correctly or incorrectly.*

The original P2022 paper obtained a Bayes factor of about 60 favoring a two-successful-introduction model over a single-successful-introduction model. Three coding errors in those calculations were noted on the "pubpeer" post-publication review site [3] by Angus McCowan (under a pseudonym) and described by him in detail on an arXiv paper. [2] The current online version of P2022 now includes corrections for those three errors. The corrections reduced the likelihood ratio favoring two introductions from ~60 in the original version to ~4.3 in the current version.

## Overview of the Remaining Corrections.

It is obviously essential that the same selected properties of the data be used for each hypothesis when Bayesian calculations update estimates of the odds of two hypotheses by taking the ratio of the conditional probabilities of those data for the two hypotheses, $P(data|I_2)/P(data|I_1)$. The conditional probability of more-detailed properties of the data will be smaller than the conditional probability of less-detailed properties.

For example, if one were to update the odds for deciding which of two suspects committed a burglary from which a blue Toyota was seen leaving the crime scene, it would not be correct to use the ratio P(drives car|suspect 2)/P(drives blue Toyota|suspect 1). That comparison would

incorrectly favor the "suspect 2" hypothesis since only a fraction of car drivers happen to drive blue Toyotas. That fundamental error in logic is precisely analogous to the core error of P2022. The main issues that I shall discuss concern corrections to the $I_2$ likelihood used by P2022 when it is estimated using the same observed properties as were used for $I_1$. In other words, I use the analog of the full "blue Toyota" condition for each hypothesis.

P2022 calculated $P(\tau|I_1)$ from 1100 simulations of single introductions with some favored parameters. (These parameters were a nominal doubling time of 3.47 days, an ascertainment probability of 0.15, and 100 descendant branches used as the minimum for a polytomy.) Only 34 simulations met the criteria used for $\tau$. That gives a point estimate $P(\tau|I_1) = 0.031$, with a standard 95% confidence interval of [0.022, 0.043].

*P2022 did not model or simulate the two-introduction $I_2$ account.* Instead, P2022 calculated $P(\tau|I_2)$ by using only the first two of the four conditions I listed that were required of the $I_1$ simulations, whether an introduction produces a large enough basal polytomy and was compatible with the probable MRCA. P2022 denotes the minimum-size basal polytomy criterion for a single introduction as "$\tau_P$". Of P2022's 1100 $I_1$ simulations, 523 meet that $\tau_P$ criterion. P2022 describes the process of extrapolating to $I_2$ as follows:

> "We assume each introduction is independent, allowing us to generalize this probability to $P(\tau|I_n)$. For example, $P(\tau=\tau_P|I_n=I_1) = 0.475$ and $P(\tau=(\tau_P, \tau_P)|I_n = I_2) = P(\tau=\tau_P|I_n=I_1)^2 = 0.226$."

This formula makes it unambiguously clear that the two introductions are meant to each consist of a single introduction of the type described by the simulations, since the numerical $\tau_P$ is directly taken from those single-introduction simulations. The likelihood ratio P2022 obtains is then proportional to $(P(\tau_P|I_1))^2/P(\tau|I_1)$. Including the complicated MRCA-related condition, the second on my list of four, reduces the P2022 estimate of $P(\tau|I_2))/P(\tau|I_1)$ from $P(\tau_P|I_1))^2/P(\tau|I_1)$ by a factor very close to 0.6, giving 4.3 for the main estimate of the Bayes factor favoring $I_2$.

The central problem with this method of estimating $P(\tau|I_2)$, as first noted by McCowan [2], is that it does not include the conditions "comprising between 30 and 70% of taxa, two mutations from the root with no intermediate genomes". These are the third and fourth conditions on my list.

Since the two introductions are described as independent, one can obtain a limit on the distribution of the ratio of the sizes of the two resulting lineages simply from the range of sizes generated by the P2022 model of single introductions. The narrowest distribution and thus the highest probability of meeting the relative size condition is obtained if the two introductions are simultaneous, i.e. with neither having a head start.

The condition that the two lineage sequence roots be separated by two nucleotides, i.e. D = 2, is indirectly related to the MRCA-related condition. P2022 shows that the estimated probabilities of different MRCAs depend strongly on how the estimation is made, which makes it somewhat complicated to extrapolate this criterion to $I_2$. To be conservative, I shall ignore that $I_1$-favoring MRCA-dependent factor since it is not fully independent of the D=2 condition. Instead, I shall just use the simple observed sequence difference D=2. An approximate upper limit on $P(D=2|I_2)$ can be derived from some minimal assumptions about the prior evolution.

Using the D=2 constraint rather than the more complicated constraint on the relation of the observed sequences to the hypothetical MRCAs allows a major simplification in the analysis. For $I_2$ meeting the size ratio condition depends only on the *difference* in the times of the two introductions but meeting the D=2 condition depends only on the *sum* of the two times after their MRCA. That decouples the two constraints, allowing each to be treated separately by simple methods.

## Fixing the Size Ratio Constraint

For $I_1$, P2022 required that the fraction of the sequences in the smaller lineage be at least 30%, i.e. $|\ln(SR)| \le 0.85$. Since P2022 "assume each introduction is independent" the growth of each

clade starting from its introduction into humans is independent and thus can be described by the model and simulations used for $I_1$. (The P2022 calculation of the probability that both lineages meet the $\tau_P$ condition confirms that this independence was assumed in its calculations for $I_2$.) Although P2022 discusses the probability under $I_2$ that the two introductions might have occurred at detectably different times one can find the maximum probability of getting similar-sized lineages from each introduction by not giving either one a head-start, i.e. assuming that the introductions are simultaneous. The size distribution is broad enough for that simultaneity to also maximize the probability density near the observed SR value.

The imbalanced use of the clade size ratio condition can then be fixed by examining the relative size distribution of the collection of pairs of single-introduction results generated directly by the phylogenetic model and simulations of P2022. This correction factor to $P(\tau|I_2)$ is the fraction of such pairs that have close enough sizes after a fixed time near when the simulation reaches 50,000 net cases, the point at which sampling for the statistical tests stops. Determining that fraction requires looking at the collection of pairs of single introductions for which the resulting sequences meet the other P2022 criteria.

Before giving a more precise answer extracted from the P2022 supplementary files, it may be instructive to show how a reader might obtain an approximate value directly from the published results. The key step in that process uses that simulations that take longer to reach a fixed size will be smaller after a fixed time. The necessary information about the distribution of times is contained in the next figure.

By the time over 10,000 sequences are generated in a simulation the growth has become close to uniformly exponential with a doubling time a bit less than the nominal value for these simulations, 3.47 days, as illustrated in Fig. S22D. That allows the distribution of times to reach some size to be converted to distributions of logarithmic sizes at a fixed time by a simple conversion factor.

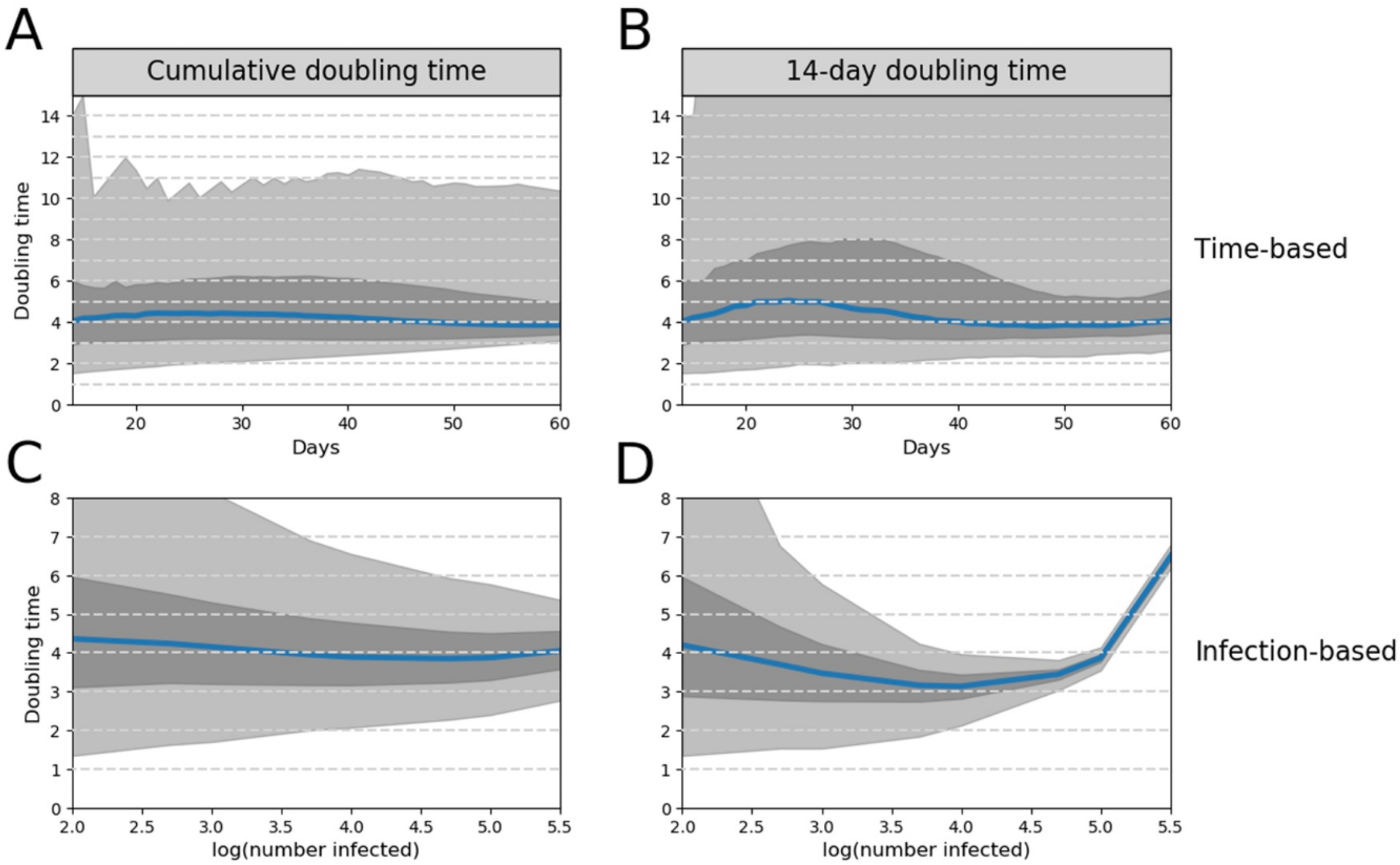

**Figure S22. Inferred doubling times of simulated epidemics.** Inferred doubling times of the 1100 primary simulations. (**A**) Cumulative doubling time since the start of the simulation. (**B**) 14-day doubling time from day 14 until the end of the simulation, with cumulative doubling time reported prior to day 14. (**C**) cumulative doubling time once a certain number of individuals are infected (*e.g.,* the cumulative doubling time at the 100th infection). (**D**) 14-day doubling time once a certain number of individuals are infected, with cumulative doubling time reported if that number of infections occurred before day 14 in the simulation. The center blue line represents the median doubling time across the simulations. Darker and lighter shading represent the 50% and 95% HDI, respectively.

Fig. 2. Fig. S22 of the Supplement to P2022, used here under the Creative Commons license https://creativecommons.org/licenses/by/4.0/.

Since at any particular time there is a substantial spread in the sizes of the those $I_1$ results only a relatively small fraction of the pairs are close enough in size to meet the condition $|\ln(SR)| \leq 0.85$. Here is a clarifying example. Say that one particular simulation happens to have taken the median length of time to reach ~32,000 cases, i.e. 15 doublings. According to Fig. S22C, its cumulative doubling time, including the highly stochastic initial stages, is ~3.9 days, so those 15 doublings took 3.9*15= 58.5 days. A simulation at the 25$^{th}$ percentile has a cumulative doubling time of about 4.6 days, so it would take 69 days to reach that many cases. It is then 10.5 days behind. With a current (not cumulative) doubling time of 3.5 days it would be a factor of $2^{10.5/3.5}$ = 8 short of the number of cases for the median simulation. Therefore it would not be close enough in size for the pair to be counted as meeting the $\tau$ criteria. This one example already shows that even for a simulation with a median rate substantially less than half of the other simulations are close enough in size to meet the relative size criterion.

We can now take a more systematic look at the distribution to see about how many pairs are close enough in size. P2022's Fig. S22C shows the width of that time distribution near the end of the simulations, e.g. at $10^{4.5}$ sequences, corresponding to 15 doubling times. At that point the median cumulative doubling time, including all the early more stochastic steps, is 3.9 days. Fig. S22C shows the intervals containing 50% and 95% of the cumulative doubling times for the collection of simulations, giving about (3.2,4.7) and (2.2,6.0) days, respectively. These would be consistent with Gaussian approximations with standard deviations of 1.13 days and 0.97 days, respectively. The form of the tails of the distribution will have negligible effect on the fraction of pairs that are close enough in size. For our approximate purposes here a Gaussian distribution with width 1.05 days should be adequate.

Fig. S22D shows the current (within 14 days) doubling times. In the relevant size range, the median doubling time is ~3.25 days. (That time is less than the nominal doubling time, presumably because the cases tend to accumulate in the more-connected nodes until, later, those nodes saturate, an effect also seen in Fig. S22D.) Thus we may approximate the distribution $\rho$(ln(size)) toward the end of the simulation by using a Gaussian with standard deviation 15*1.05 days*ln(2)/3.25 days = 3.36.

The distribution then has the approximate form:

$$\rho(\ln(size)) = (\frac{1}{3.36})(2\pi)^{-0.5}e^{-(\frac{0.5}{11.3})(\ln(size))^2}$$

If all the $I_1$ simulations contributed to the results that met the other criteria, then the distribution of ln(SR) would just be a Gaussian with width $3.36*2^{0.5} = 4.75$ since the sizes of the clades descended from the two introductions would vary independently. Then only 14% would have $|\ln(SR)| \leq 0.85$, i.e. have $|z| \leq 0.85/4.75 = 0.18$ on a normal distribution.

The subset of simulations meeting the requirement of having a minimum polytomy size will tend to have a narrower distribution of sizes since the number of taxa and the basal polytomy's number of branches are likely to be correlated. One may make an extreme allowance for such an effect by eliminating the smaller half of the size distribution. Taking the integral numerically, 26% of the remaining pairs would then fall within the required size ratio range. A cautious estimate of the correction for the clade-size size constraint would then be a factor in the range 0.14 to 0.26, with a geometric mean of 0.19.

Now we can look at a more precise number not readily available to a casual reader. I will use the same essential technique— the nearly linear relation between the time to reach a large case number (50,000) and the log of the number of cases at fixed times a little before. J. Pekar has provided guidance on how to obtain the relevant files. Angus McCowan has supplied a script (available in the Supplemental material to this paper) to extract the times to reach 50,000 cases for just the subset of simulations that meet the $\tau_P$ constraint rather than the times for the larger set represented in P2022's Fig. S22.

One may get a feel for the distribution of times to reach 35,000 cases for the 523 simulations using the favored parameters that met the $\tau_P$ criterion from a histogram of the pairwise differences in those times, symmetrical about zero since the order of the simulations is arbitrary. Self-pairs are not included since they do not represent independent simulations.

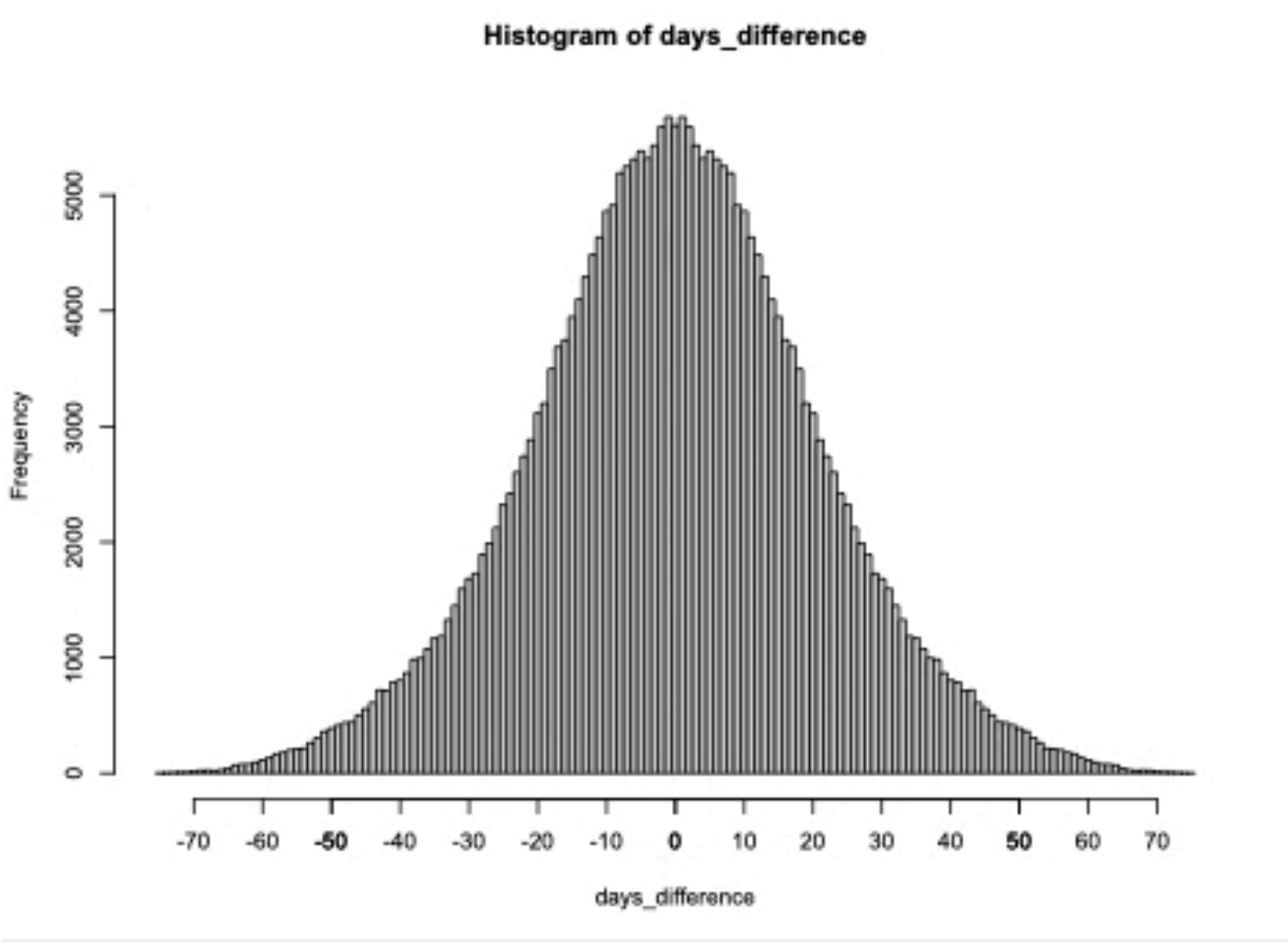

Fig. 3 The histogram of differences of the times to reach 35,000 cases for pairs of P2022's 523 simulation runs that met the $\tau_P$ criterion.

With a 3.25 day doubling time, a 4-day lag corresponds to a factor of 2.35 in size, almost fitting the window of up to a factor of 7/3= 2.33 used as a condition on $I_1$. Of the 523 simulations using the favored parameters that meet the $\tau_P$ criterion there are 19499 pairs (excluding self-pairs, which do not represent independent simulations) whose completion times differ by three days or less. Essentially all of these pairs should meet the size constraint. In addition there are 5,319 pairs with a four-day difference. Close to half of those should meet the size constraint. Since there are 523*261=136,503 non-self pairs, the fraction that meets the size constraint is about 22158.5/136,503 = 0.162.

This estimate is slightly larger than the estimate from the full set of Fig. S22C and is smaller than the upper bound obtained from looking at half the distribution of Fig. S22C. A reader looking at the published figure could estimate the correction factor rather closely. The 0.162 correction factor applied directly to the P2022 Bayes factor (which includes the MRCA-related condition) would reduce it from 4.3 to 0.70, eliminating any tendency for the analysis to favor $I_2$.

## Fixing the Sequence Difference Constraint

Rather than use the simple observed D=2 constraint P2022 uses two different ways of estimating the probabilities of different root sequences for $I_1$, then weights the different simulated topologies by their consistency with those roots. Although the Bayes factors for these different ways of assigning the probable MRCA's end up being almost identical, their probabilities of the different MRCAs scarcely overlap. Bloom has pointed out that the evident clustering of cases and non-random sampling of clusters reduces the reliability of any root assignment based on early sequences, and the assignment based on related viruses is not uniquely determined.[4]

Going forward, for simplicity and conservatism I shall drop the $I_1$-favoring MRCA-related factor since it overlaps the D=2 condition. Including the size-ratio factor of 0.162 with the P2022 $\tau_P$ condition lowers the probability that two simulations would give adequate polytomies and have close enough sizes to 0.037.

P2022 describes the simulations meeting the criteria for $I_1$ as having the sequence difference between the two clades as D =2 nucleotides. Without specified priors $I_2$ can have arbitrarily large D because there are no relevant limits on the pre-introduction sequence diversity. Thus, unlike for the relative size condition, the distinction between the proper Bayesian requirement D=2 and an improper one-sided extension is crucial. Since $I_1$ tends to give small D a one-sided extension of the observed D=2 condition to D < 3 would give a completely different likelihood ratio than would the opposite one-sided extension to D > 1.

For $I_2$, as for $I_1$, the observed result D=2 is not especially easy to meet. One can get a qualitative feel for the problem by visual inspection of P2022's Fig. S30, above. Two sequences happen to spill over from the hypothetical host to humans. Focusing on the sequences derived from the MRCA, there are 28 possible introductory pairs, included self-pairs, which there is no reason to exclude. Sequence differences correspond to horizontal lengths on the paths connecting two sequences. Of the 28 pairs, only 6 have D=2, i.e. only 21%. We need, however, a quantitative analysis of $P(D=2|I_2)$ rather than counts from an illustrative sketch.

Each introduction is of a sequence descended from their MRCA, whatever that might happen to be. There are a very large number of possible mutations (~90,000) that could have occurred but one finds only the small observed value D=2 between the two big clades. Although the probabilities of individual mutations are not all equal, P2022 shows hundreds of mutations detected, none of which are at all likely to occur in any single line of descent on the time scale of weeks. Thus the probability for each mutation on a path from the MRCA to an introduction must be quite small. There is no indication that the two mutations separating A and B are strongly linked to each other by some constraint since no such linkage was used in the P2022 forward-going simulations and since in actual cases intermediate sequences appeared soon[5].

The sequence difference $D_0$ between $I_2$'s two introductions thus comes from the sum of a large number of low-probability events and therefore should have a probability distribution very close to the Poisson limit. Its probabilities are then fully characterized by its expectation value $E(D_0)$. $E(D_0)$ depends on the sum of the times from the MRCA to the two introductions multiplied by the mutation rate in the hypothetical prior host, but we do not need to consider those factors separately.

Mutations after the introduction but before the detected clade root can also contribute to the net D. These, like $D_0$, will have a distribution of values around their mean and thus cannot be fine-tuned to produce D=2. For our purposes, the Poisson distribution approximation for D should suffice. The minor effects of post-introduction mutations, slightly lowering $P(D=2|I_2)$, are discussed in the second Supplement.

The maximum possible Poisson $P(D=2|I_2)$ is found for E(D)=2, giving $P(D=2|I_2) \leq 2/e^2 = 0.27$. Reaching this probability, however, requires fine-tuning E(D) to precisely 2, without any prior justification. We may consider instead a plausible prior distribution for E(D), $\rho_0(E(D))$, and obtain a posterior distribution $\rho(E(D))$ by conditioning on the observed D=2, and then calculate P(D=2) by marginalizing over $\rho(E(D))$. In effect, this procedure still allows some post-hoc adjustment of the $I_2$ model to agree with observations but it does not allow unrealistic fine-tuning.

For brevity, I shall denote E(D)=x. One plausible prior $\rho_0(x)$ would be simply uniform up to truncation at some irrelevant large value. One then obtains $\rho(x)= x^2e^{-x}/2$. Marginalizing over that gives P(D=2)=3/16, close to the accidental value found from Fig. S30. This factor is a bit smaller than the fine-tuned $2/e^2$. Another plausible prior would be uniform in ln(x), again truncated at irrelevant extreme values. It gives $\rho(x)= xe^{-x}$. Coincidentally, marginalizing over that also gives P(D=2)=3/16.

For a general power-law prior $\rho_0(x)$ proportional to $x^{-\alpha}$ with $\alpha$ in the range for which divergences are not problematic $P(D=2) =(4-\alpha)(3-\alpha)/2^{(6-\alpha)}$ . The maximum is obtained for

$\alpha = (7*\ln(2)-2-(4+\ln(2)^2)^{0.5}/(2*\ln(2)) = 0.53$

which gives P(D=2) = 0.193.

## Bottom line

Incorporating the size and sequence difference constraints for $I_2$ under conditions chosen to favor $I_2$ gives a corrected $P(\tau|I_2) \leq 0.475^2*0.162*0.193 = \sim 0.0071$. The approximate resulting likelihood ratio is $0.031/0.0071 \geq 4.4$ favoring $I_1$ over $I_2$. Within the P2022 approach, using balanced observational features thus gives a likelihood ratio favoring the single-introduction picture over the two-introduction picture even though the features used were chosen because they superficially suggested two introductions.

## Further Minor Points within the P2022 Model

### Adjustments to $P(\tau|I_1)$ and $P(\tau_P|I_1)$

McCowan [2] has discussed further minor coding fixes required to bring the code into alignment with the algorithm described in the P2022 text as well as refinements of the precision obtained by using 100 times as many simulations. Those steps can result in further small changes in $P(\tau|I_2)/P(\tau|I_1)$.

### Minimum polytomy size for pairs

The P2022 algorithm uses a minimum polytomy size to evaluate if the $\tau_P$ condition has been met in each single introduction by the time the simulation reaches 50,000 cases, the number used for the statistical tests. When two introductions are combined in $I_2$ the statistical tests should also use the first 50,000 cases. Therefore the number from each introduction would be less than 50,000. That reduction could slightly reduce $P(\tau_P)$ and thus slightly reduce $P(\tau|I_2)$.

### Number of Data Points

The main estimate used in P2022 is based on 7500 sequences drawn from simulated sets of 50,000, i.e. with each sequence having a probability of 0.15 of being detected. The actual data used, however, have only 787 sequences. Thus there is an inconsistency between the data used and the simulation output sampling used. Since the core observation on which the $I_2$ hypothesis was based was the lack of intermediates between lineages A and B, use of a more realistic sparse sampling might increase $P(\tau|I_1)$.

### The MRCA Factor

We can now look back at P2022's MRCA-related factor of ~1.7 favoring $I_1$ that I've omitted from this analysis to try to understand it qualitatively in the framework used here. For $I_1$ almost all of P2022's simulations that meet the other criteria have the MRCA at the root of one of the clades. The possibility that neither of the roots is the MRCA is assigned a low probability. For $I_2$, however, when the two introductions are simultaneous in the approximation $D=D_0$ the Poisson statistics would imply that half the $I_2$ pairs meeting the other criteria would not have the MRCA at either root. The probability of having the MRCA at a root could be raised, in that approximation, toward 1.0 by having the post-MRCA time for one introduction be close to zero. That difference in times, however, would reduce the probability of the two sizes being close enough. Thus using the MRCA criterion complicates the analysis by coupling the parameters describing the sum of the post-MRCA times and their difference. The case for believing that one of the roots is the MRCA is not strong enough to justify exploring those complications here. McCowan's more detailed analysis[2] does include this factor.

## Other Statistical Issues

My core conclusion is that the claim of P2022 to have shown that two introductions were more likely than one was based on errors in Bayesian reasoning in applying its model to its data using its simulations. Correcting the explicit errors reverses the conclusion.

Some other more familiar statistical issues may also affect the substantive conclusion. Realistic odds require not only correct application of a particular model to a limited data set but also some evaluation of the limitations of the model and of the data. Here I discuss the data limitations informally.

The form of the P2022 model may bias the results because there are non-random selection effects on the path from the full set of sequences to the small subset available for analysis. Bloom [4] has described evidence that different locations were sampled at much different rates. Likewise, although P2022 assumes time-independent rates of sampling, at the early stage the detection rate was particularly low [6]. These uneven sampling issues were also noted by S. Zhao et al. who wrote, "In the coalescent process of their simulations, they assumed that viruses spread and evolve without population structure, which is inconsistent with viral epidemic processes with extensive clustered infections, founder effects, and sampling bias." [7] The unevenness of the sampling in both location and time widens the path for "no intermediate genomes" to be found in the data set even if they had existed. If so, one might expect some reliable intermediates to show up soon afterwards, as Lv et al. [5] subsequently found in a more complete data set. Using more realistic descriptions of the early sampling could easily undercut the main rationale for considering $I_2$, which was that if $I_1$ were true it was unlikely that early intermediates would be missed.

## Conclusion

The P2022 Bayesian analysis of the number of successful introductions had major errors in basic Bayesian techniques. Correcting these explicit errors reverses the direction of the conclusion. Based on the P2022 model and data a single introduction is the more likely interpretation, as Lv et al. [5] concluded from their interpretation of more complete data. The corrected Bayesian analysis of the P2022 data, however, still allows a chance for two successful introductions to have occurred.

## Acknowledgements

I am greatly indebted to Angus McCowan for noticing the imbalanced conditions re-analyzed here and for stimulating technical discussions. I'm even more grateful to him for extracting the data on simulation sizes, a task beyond my capabilities. I hope that he will publish a more comprehensive analysis without some of the approximations that I had to use here to extract estimates from the published paper.

## Supplements

**1) This Supplement includes the Python script for extracting time-to-completion data, the list of relevant times, and the R program for finding how many were close enough.**

**Here's the Python script to get the CSV file of the completion time of simulations that meet the polytomy requirement:**

```
#!/usr/bin/env python3
# coding: utf-8

# #### Identify the runs with basal polytomies
# Download the clade analysis results from https://github.com/sars-cov-2-origins/multi-introduction/raw/refs/heads/main/FAVITES-COVID-Lite/cumulative_results/FAVITES_results.zip.
# Unzip in the same directory as this script.
import os
runs_with_polytomies = []
for i in range(1, 1101):
    fpath = os.path.join("clade_analyses_CC", f"{i:04d}_clade_analysis_CC_polytomy.txt")
    with open(fpath) as f:
        n = sum(1 for _ in f) # count lines in the file
    if n >= 100: # each line represents a subclade or basal lineage, i.e. a branch of the
basal polytomy
        runs_with_polytomies.append(i) # record if 100 branches or more
print(f"Number of runs with basal polytomies: {len(runs_with_polytomies)}")
print(f"Frequency of basal polytomies: {100*len(runs_with_polytomies)/1100:.1f}%")

# #### Get day of 50kth infection for the runs with basal polytomies
# Download the simulation results from https://github.com/sars-cov-2-origins/multi-introduction/raw/refs/heads/main/FAVITES-COVID-Lite/cumulative_results/FAVITES_GEMF_dict.pickle.zip.
# Unzip in the same directory as this script.
import pickle
with open("FAVITES_GEMF_dict.pickle", "rb") as f:
    gemf = pickle.load(f)
results = {}
for run in runs_with_polytomies:
    run_id = f"{run:04d}"
    results[run_id] = 100 # default to end of simulation
    for day in range(101):  # 0..100 days of simulation
        if gemf[run_id][day]["S"] < 4950000: # 'S' is the susceptible compartment and starts
from 5,000,000
```

```
            results[run_id] = day
            break

# #### Write to csv
import csv
with open("day_of_50k.csv", "w", newline="") as f:
    writer = csv.writer(f)
    writer.writerow(["run_id", "day_of_50kth"])
    for run_id, day in results.items():
        writer.writerow([run_id, day])
```

**Here's the R program and the list of completion dates, obtained by converting the .csv output to a comma-delimited list and converting "100" to the minimum "101" for runs that did not reach 50k. The last part of the program produces the histogram.**

```
> days<-
c(37,51,63,65,49,42,45,58,64,73,71,55,36,38,62,75,47,46,46,74,35,71,51,52,55,54,59,
87,31,43,49,49,61,85,38,66,46,61,39,54,32,43,47,54,43,46,46,43,42,42,42,32,36,68,5
9,49,61,64,59,46,36,58,80,35,71,41,66,47,43,88,58,42,49,48,36,84,53,53,59,49,52,49,
61,57,77,53,37,49,47,67,69,95,39,33,52,89,34,43,41,80,52,51,61,71,53,51,61,55,79,5
6,44,44,58,73,58,66,49,56,90,50,61,62,53,69,67,48,51,55,43,68,46,35,61,47,48,40,70,
77,60,52,38,56,92,51,78,87,79,87,51,58,46,68,50,62,48,38,63,52,71,63,58,46,49,56,4
9,45,89,62,50,36,53,99,64,43,71,59,38,33,72,47,57,31,34,29,54,63,41,71,49,59,35,26,
55,56,73,59,44,43,43,68,50,48,43,42,42,90,41,57,48,74,58,84,60,53,60,61,43,100,44,
48,47,48,57,40,37,64,74,59,44,40,40,62,50,54,28,45,54,41,42,57,46,38,57,52,30,42,7
8,50,49,36,48,51,41,44,32,59,37,51,66,67,91,53,61,60,51,53,57,59,59,37,45,56,40,63,
34,52,39,35,81,55,57,57,94,62,67,48,47,39,61,49,44,51,49,36,46,39,54,63,48,60,73,5
4,59,65,77,75,63,60,49,48,32,41,74,56,79,56,67,51,58,42,59,43,49,44,41,51,64,40,54,
51,75,28,72,54,54,91,69,34,43,37,64,57,57,36,54,48,75,51,51,60,43,66,54,73,55,50,5
9,86,49,47,45,40,50,53,50,82,32,75,68,51,47,93,63,62,52,39,42,77,50,53,44,45,64,45,
100,70,48,52,101,54,66,70,37,31,67,47,60,51,46,48,73,66,62,40,31,79,74,60,44,44,48
,44,48,40,66,43,58,41,44,69,37,53,49,68,49,47,60,66,57,59,60,39,67,30,67,92,59,55,4
9,44,36,49,60,49,59,46,45,49,38,53,53,42,50,49,59,36,27,42,55,37,37,68,50,49,51,57,
92,74,76,81,32,31,94,57,35,63,56,80,41,30,52,39,29,50,41,42,42,50,87,45,69,41,32,7
2,76,28,51,41,57,51,63,92,81,59,50,43,31,70,65,58,52,67,55,66,84,43,42,46,36,52,50,
71,74)
> L<- length(days)
> count <- 0
> for (i in 1:(L-1)) {{for(j in (i+1):L)  {if (abs(days[i]-days[j]) <= 3) {count <- count + 1}}}}
>count
[1] 19499
```

```
>for (i in 1:(L-1)) {{for(j in (i+1):L)  {if (abs(days[i]-days[j]) == 4) {count <- count + 0.5}}}}
> count
[1] 22158.5
> 2*count/(L*(L-1))
[1] 0.1623298

> days_difference <- 1:(L*(L-1)/2)
> zz <- 0
> for (i in 1:(L-1)) {
    + for(j in (i+1):L)  {zz<-zz+1
    + days_difference [zz] <- (days[i]-days[j]) }}
> for (i in 1:( L*(L-1)/2)) { days_difference [(L*(L-1)/2)+i] <-  -days_difference [i]}
> hist(days_difference, breaks= -75.5:75.5)
> axis(side = 1, at = c(-7:7)*10)
```

This gives 22,158.5/(1099*550) = 0.037 for P(polytomies & size ratio|$I_2$), not yet considering D=2. The factor 0.162 reduces the P2022 Bayes factor to 4.3*0.162= 0.70, not yet considering D=2.

2) This supplement looks at the effects of the post-introduction pre-detected-clade-root mutations on $P(D=2|I_2)$. This extra mutation number is not necessarily describable by Poisson statistics because the clade roots are not randomly selected but are chosen by the P2022 algorithm in a way that depends on their difference from other sequences.

The resulting change in $P(D=2|I_2)$ turns out to be quite small. Inclusion of these post-introduction pre-root mutations may be more important in evaluating probabilities of more complicated multi-introduction hypotheses since this post-introduction diversity is inherent in the P2022 single-spill model and cannot be altered by adding more complications to the hypothesized pre-introduction stage.

We can denote the probabilities of the detected root being (0,1,2) nucleotides different from its introductory sequence by $(p_0,p_1,p_2)$. To obtain D=2 one needs that the sum of the Poisson pre-introductory $D_0$ and the two post-introductory changes be two. Adding the probabilities for each such path, again calling $E(D_0)=x$, gives:

$$P(D=2|I_2) = e^{-x}\,(2\,p_0p_2+p_1^2+2p_0p_1x+0.5\,p_0^2x^2).$$

The maximum is obtained when x is fine-tuned to

$$x = (\,p_0-2\,p_1+(p_0^2+2p_1^2-4p_0p_2)^{0.5})/p_0.$$

McCowan (personal communication) has extracted the post-introduction diversity distribution from the 1100 relevant P2022 simulations. The simulations give $(p_0,p_1,p_2)$.=(0.589, 0.235, 0.085) for the probabilities of 0,1, or 2 mutations, the most important values. Then the fine-tuned maximum $P(D=2|I_2)$ occurs for x=1.06 which gives $P(D=2|I_2)$= 0.22, less than the 0.27 obtained for the fine-tuned maximum when the post-introduction diversity was ignored. Considering power-law priors for x, just as we did without including the post-introduction diversity, reduces the maximum to 0.176, slightly smaller than the value obtained before. Thus inclusion of the post-introduction diversity makes little difference for $P(D=2|I_2)$.

Recently, as the extreme problems with the P2022 analysis have become clearer, informal suggestions have arisen for more complicated stories of multiple successful introductions. Multiple introductions of a single sequence can raise the probability that the lineage with that sequence will end up with a size not too far below the median. For each such indistinguishable introduction, however, there is a probability of $1-(p_0^2 + 2p_0p_1) = \sim 0.38$ of getting an additional distinguishable clade root 2 nt or more different from the others. Any attempt to restore the probability of multiple successful introductions by replacing the P2022 approach with a hypothesis of more than two successful introductions would need to account not only for the extra adjustable parameters required but also for the effects of such extra distinguishable clades on the topology.